\documentclass[a4paper,fleqn,usenatbib]{mnras}

\usepackage{newtxtext,newtxmath}

\usepackage[T1]{fontenc}
\usepackage{ae,aecompl}

\usepackage{graphicx}	
\usepackage{amsmath}	
\usepackage{amssymb}	

\newcommand{\lta}{\mathrel{\hbox{\raise 0.6 ex \hbox{$<$}\kern
                   -1.8 ex\lower .5 ex\hbox{$\sim$}}}}                
\newcommand{\gta}{\mathrel{\hbox{\raise 0.6 ex \hbox{$>$}\kern
                   -1.7 ex\lower .5 ex\hbox{$\sim$}}}}
                   
\DeclareMathAlphabet{\mathsc}{T1}{cmr}{m}{sc}

\title[Non-axisymmetric 3D element stratifications]{Three-dimensional abundance
  distributions in ApBp star atmospheres: non-axisymmetric magnetic geometry}

\author[G. Alecian \& M.J. Stift]
{
G.~Alecian$^{1}$\thanks{E-mail:georges.alecian@obspm.fr},
M.J.~Stift$^{2,3}$
\\
$^{1}$LUTH, Observatoire de Paris, PSL Research University, CNRS, Universit{\'e} Paris
    Diderot, Sorbonne Paris Cit{\'e}, \\5 place Jules Janssen, F-92190 Meudon, France\\
$^{2}$Armagh Observatory, College Hill, Armagh BT61 9DG. Northern Ireland\\
$^{3}$Kuffner-Sternwarte, Johann Staud-Strasse 10, A-1160 Wien
}

\date{Accepted XXX. Received YYY; in original form ZZZ}

\pubyear{2016}

\begin{document}
\label{firstpage}
\pagerange{\pageref{firstpage}--\pageref{lastpage}}
\maketitle

\begin{abstract}
Numerical models for the atmospheres of magnetic ApBp stars have in the past dealt only with centred dipole magnetic field geometries. These models include atomic diffusion that stratifies the abundances of metals according to the local magnetic field strength and the direction with respect to the surface normal. The magnetic variations with rotational phase of most well observed stars however reveal that this assumption is far too simplistic. In this work we establish for the first time a three-dimensional (3D) model with abundance stratifications arising from atomic diffusion of 16 metals, adopting a non-axisymmetric magnetic field geometry inspired by the configuration derived for a real ApBp star. We find that the chemical elements are distributed in complex patterns in all three dimensions, far from the simple rings that have been proposed as the dominant abundance structures from calculations that assume a perfectly centred dipolar magnetic geometry.

\end{abstract}

\begin{keywords}
{
atomic diffusion -- stars: abundances -- stars: chemically peculiar --
magnetic fields, stars : magnetic fields 
}
\end{keywords}



\section{Introduction}

Our current understanding of magnetic ApBp star atmospheres -- in which the vertical and horizontal distributions of chemical abundances are not homogeneous -- has recently been discussed in detail by \citet{AlecianAl2015} within the framework of atomic diffusion theory. Numerical results on diffusion in ApBp star atmospheres have been applied to idealised stars, invariably under the assumption of a magnetic geometry based on a simple centred dipole \citep{AlecianAlSt2010}. It has been shown that when the stratification process has reached equilibrium -- i.e. the particle flux is zero -- many metals (but not all of them) are expected to exhibit large overabundances around the ``magnetic equator'' (where the field lines are tangent to the surface, $90\pm{5}\degr$), often in layers above $\log\tau\approx{-2.0}$. The same models predict mild abundance anomalies in the polar regions, and horizontally homogeneous stratifications of elements in deep layers where atomic diffusion is no longer sensitive to the magnetic field. Note that for atmospheres of non-magnetic ApBp stars (mainly HgMn stars), abundance stratifications depend only on depth; they are therefore uniformly distributed over the surface \citep{LeBlancLeMoHuetal2009l, AlecianAlSt2010}, except for those cases where very weak magnetic fields (of a few Gauss) exist. For chemical elements with low cosmic abundance, the latter will help atomic diffusion to form field-dependent overabundance clouds at very high altitude.

According to the numerical models mentioned above, abundance rings or belts around magnetic ApBp stars should be commonly observed. This is not the case \citep[see for instance][]{SilvesterSiKoWa2014b}. Several reasons may be invoked to explain the lack of agreement between the surface abundance distributions of a given star obtained through Zeeman Doppler mapping (ZDM), and recent numerical diffusion models. For instance, will the application of present-day ZDM algorithms to spectropolarimetric data always (or ever) lead to the recovery of the predicted abundances rings? This problem has recently been addressed by \citet{StiftLeo2016a}. On the other hand, \citet[][Sec. 3]{AlecianAl2015} has listed a number of improvements that have yet to be included in numerical models in order to fully describe the build-up of abundance stratifications in individual stars. Among these improvements we find the need to go beyond the strictly dipolar geometry that has been used until now. This is the main purpose of our present work.

We have computed a grid of 81 plane-parallel model atmospheres ($T_{\rm eff}=10\,000$\,K, $\log{g}=4.0$), adopting various magnetic field strengths and orientations; the field-dependent equilibrium abundance stratifications result from the simultaneous atomic diffusion of 16 metals (Sec.~\ref{sec:modeldistribgrid}). With newly developed tools we are now in a position to establish from this grid of models the 3D distribution of any of the 16 elements for a given magnetic geometry (Sec.~\ref{sec:modeldistrib3D}). In a final step, we have computed the distribution of two metals (Cr and Fe) for a realistic stellar magnetic configuration (Sec.~\ref{sec:magnegeo}). The results are shown and discussed in Sec.~\ref{sec:results} and Sec.~\ref{sec:disc}.

\section{Modelling three-dimensional abundance distributions}
\label{sec:modeldistrib}

Because the geometrical thickness of the atmosphere is very small compared to the stellar radius, the vertical timescales of abundance stratification processes due to atomic diffusion are much shorter (by at least 4 orders of magnitude) than the timescales for the horizontal migration of elements over the stellar surface \citep[see for instance][Sec.~5.3]{MichaudMiAlRi2015}. Since we are considering static atmospheres, abundance stratifications are due solely to vertical atomic diffusion; it is the local effect of the magnetic field orientation and strength on the vertical component of the diffusion velocity that is responsible for the horizontal abundance structures over the stellar surface. We therefore assume, as in \citet{AlecianAlSt2010}, that the surface of the star is made up of a juxtaposition of independent facets, each of them to be calculated in the approximation of a plane-parallel atmosphere. The facets differ from each other by the magnetic field strength and orientation, and by the ensuing abundance stratifications (the magnetic field is assumed to be constant with depth). There is thus a slight difference to be found between the atmospheric model used for a given facet and the model for an adjacent facet. Effective temperature and gravity are taken to be identical for all facets; potential problems and inconsistencies that could arise from this simplified 1D treatment of the local stellar atmospheres have been discussed by \citet{StiftLeo2016a}.

\subsection{Stellar atmospheres and magnetic-field dependent stratifications}
\label{sec:modeldistribgrid}

The atmospheric model for a given facet is obtained by interpolation in a grid of models (see Sec.~\ref{sec:modeldistrib3D}). The models making up this grid have been computed as described in \cite{StiftStAl2012}: they result from calculations of equilibrium stratifications with the help of the {\sc CaratStrat} code, i.e. the vertical abundance distributions are self-consistent with the atmospheric structure computed with {\sc Atlas12} (\citealt{Kurucz2005}, \citealt{Bischof2005}). The grid used in this work is composed of 81 models with $T_{\rm eff}=10\,000$\,K and $\log{g}=4.0$ and field strengths of 0, 1000, 5000, 5500, 6250, 7500, 10000, 11000, 12500, 15000, 20000\,G. Except in the 0\,G case, the models have been established for the following angles (with respect to the vertical): $0, 60, 75, 80, 83, 86, 88, 90\degr$. As in \cite{StiftStAl2012}, 16 metals are allowed to diffuse simultaneously (Mg, Al, Si, P, Ca, Ti, V, Cr, Mn, Fe, Co, Ni, Cu, Zn, Ga and Hg). This grid is 4 times larger than the one used by \citet{AlecianAlSt2010} for a given effective temperature. The parallel computations required about 43000 hours of equivalent monoprocessor time on the BullX DLC supercomputer at CINES\footnote{Centre Informatique National de l'Enseignement Sup{\'e}rieur (Montpellier, France), see: \\https://www.cines.fr/calcul/materiels/occigen/.}.

As stressed by \cite{AlecianAl2013l}, the diffusion velocity is particularly sensitive to the magnetic field orientation at and very near to the ``magnetic equator'' ($90\pm{5}\degr$). One therefore expects strong abundance contrasts in the vicinity of these horizontal fields, much less so for magnetic field angles near $0\degr$ or $180\degr$ (the magnetic poles). For this reason, the grid density increases for angles $> 75\degr$. Note that velocities are identical for $0\degr$ and $180\degr$.

\subsection{Establishing the 3D distributions of metals}
\label{sec:modeldistrib3D}

In order to obtain 3D abundance distributions over the entire atmosphere of the star, we first determine the set of facets necessary to obtain a satisfactory spatial resolution. The facets have the shapes and distribution of the area elements \citep[nicely illustrated in][]{VogtVoPeHa1987}. Field strengths and angles are specified at the centre of each area element.

In the present study, the model atmospheres are composed of $N = 72$ layers. We divide each model into a certain number of slabs. We take for instance a slab of $l$ layers (from layer number $l_1$ to $l_2$, with identical numbers for all the models) and for each chemical element we compute the average abundance (with [H] = 12) over the $l$ layers of the slab -- no weighting is applied. For our grid of 81 models we thus have a table with 81 rows and 3 columns: the field strength [Gauss], the field angle [$\degr$], and the mean abundance in the slab. In a next step we create a matrix consisting of $N_{\rm B}$ rows and $N_{\rm A}$ columns. The rows represent equally spaced field strengths, the columns equally spaced field angles, and the matrix element gives the abundance. This matrix is obtained by Voronoi interpolation applied to the 81x3 table. Here we have chosen ${N_{\rm B}} = 400$, corresponding to field strengths from $0$ to $20000$\,G in steps of $50$\,G, and ${N_{\rm A}} = 180$, corresponding to angles from $0$ to $90\degr$ in steps of $0.5\degr$. The abundance value for a given element in each slab of a given facet is then obtained by taking the matrix element closest to the field parameters of the facet (there is one 400x180 matrix per slab and per chemical element). Calculations and visualisation of the maps have been carried out with the software {\sc $^\copyright$\,Igor\,Pro\,(v7)}\footnote{See https://www.wavemetrics.com} with its built-in procedures for interpolation.

\section{The magnetic geometry}
\label{sec:magnegeo}

To be as realistic as possible, we choose the magnetic configuration of the well studied Ap star HD\,154708 whose field geometry is definitely non-axisymmetric \citep{StiftStHuLeetal2013}. The field structure can be approximated by means of the eccentric, tilted oblique rotator introduced by \citet{StiftSt1975} which is characterised by a dipole at a certain distance from the centre of the star, with its axis not going through the centre. In the case of HD\,154708 mean field modulus, mean longitudinal field and detailed intensity profiles of two Si lines are satisfactorily predicted with this particular model. HD\,154708 is by no means the only star with a clearly visible phase shift between the respective variations in field modulus and in the longitudinal field. Let us mention \citet{StiftSt1975} and \citet{StiftGoo1991} who have successfully modelled the magnetic geometries of HD\,126515 \citep{Wolff70} and of HD\,137909 \citep{Preston1970}. The star HD\,18078 \citep{MathysMaRoKuetal2016k} constitutes yet another interesting example. It is also important to realise that detailed Zeeman Doppler mapping almost invariably results in magnetic and abundance maps devoid of any simple symmetry: take for example the magnetic field map of 53\,Cam by \citet{Piskunov2008}, and the map of $\alpha^2$\,CVn by \citet{SilvesterSiKoWa2014b}. Magnetic and abundance maps of HD\,75049 \citep{Kochukhov_HD75049_2015}, of HD\,32633 \citep{Silvester_HD32633_2015}, and of HD\,125248 \citep{Rusomarov_HD125248_2016} all lack any discernible symmetries.

\begin{figure*}
\includegraphics[width=14cm]{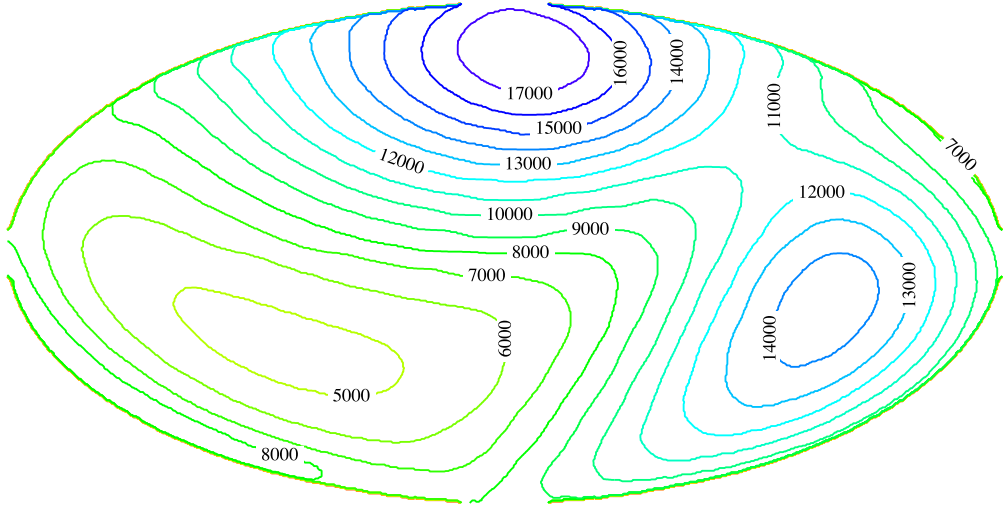}
\caption{
Contour map of the magnetic field strength used for our predicted stratifications (Hammer equal-area projection). The field strength map is based on the magnetic model for HD\,154708 \citep{StiftStHuLeetal2013}, multiplied by a constant factor (see text). Please not the complete lack of axisymmetry.}
\label{fig:Bcontour}
\end{figure*}

\begin{figure*}
\includegraphics[width=14cm]{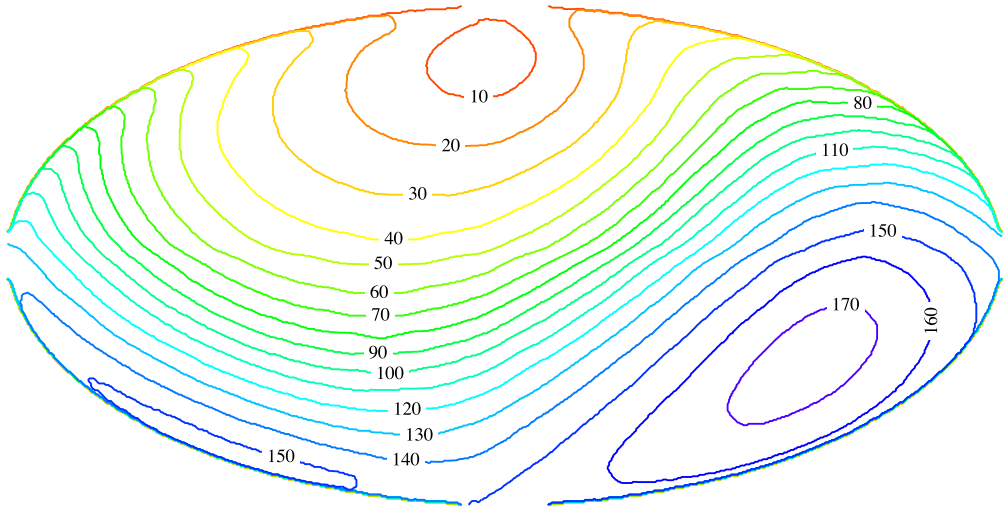}
\caption{
Same as before but showing the contour map of the angle of the magnetic field with respect to the surface vertical. The ``magnetic equator'' is defined by the $90\degr$ line. Despite the strongly non-axisymmetric geometry, the field angles largely reflect the decentred dipole.}
\label{fig:BanglesContour}
\end{figure*}

There is thus no reason to believe that the field geometry of HD\,154708 were untypical for ApBp stars. Being as yet unable to model the stratifications and their build-up in the atmospheres of specific stars \citep[see][]{AlecianAl2015}, we thus felt free to play with the geometry of HD\,154708. We have scaled the field strength to obtain abundance stratifications of sufficient contrast, while remaining inside the domain of the field parameters of our grid of models. The field strength ranges from about 5\,kG to about 17\,kG which are quite common values for a number of observed magnetic ApBp stars. The geometry underlying the results presented in Sec.~\ref{sec:results} is displayed in Fig.~\ref{fig:Bcontour} (field strength) and in Fig.~\ref{fig:BanglesContour} (field angle with respect to the vertical). The total number of facets used in this work for the description of the stellar surface is 7158.

\section{Results}
\label{sec:results}

To stay within the framework of our recently published studies, we have again chosen a main sequence atmosphere with $T_{\rm eff}=10\,000$\,K, $\log{g}=4.0$. A small part of the grid presented in Sec.~\ref{sec:modeldistribgrid} (prior to its extension to a much larger range in magnetic field parameters) was already used by \citet{AlecianAl2015}. The reader can peruse this paper for complementary descriptions/discussions of the (vertical) abundance stratifications. For the sake of conciseness, here we only present and discuss the equilibrium 3D distributions of Cr and Fe, adopting the magnetic field described in Sec.~\ref{sec:magnegeo}.

\begin{figure}
\includegraphics[width=7.5cm]{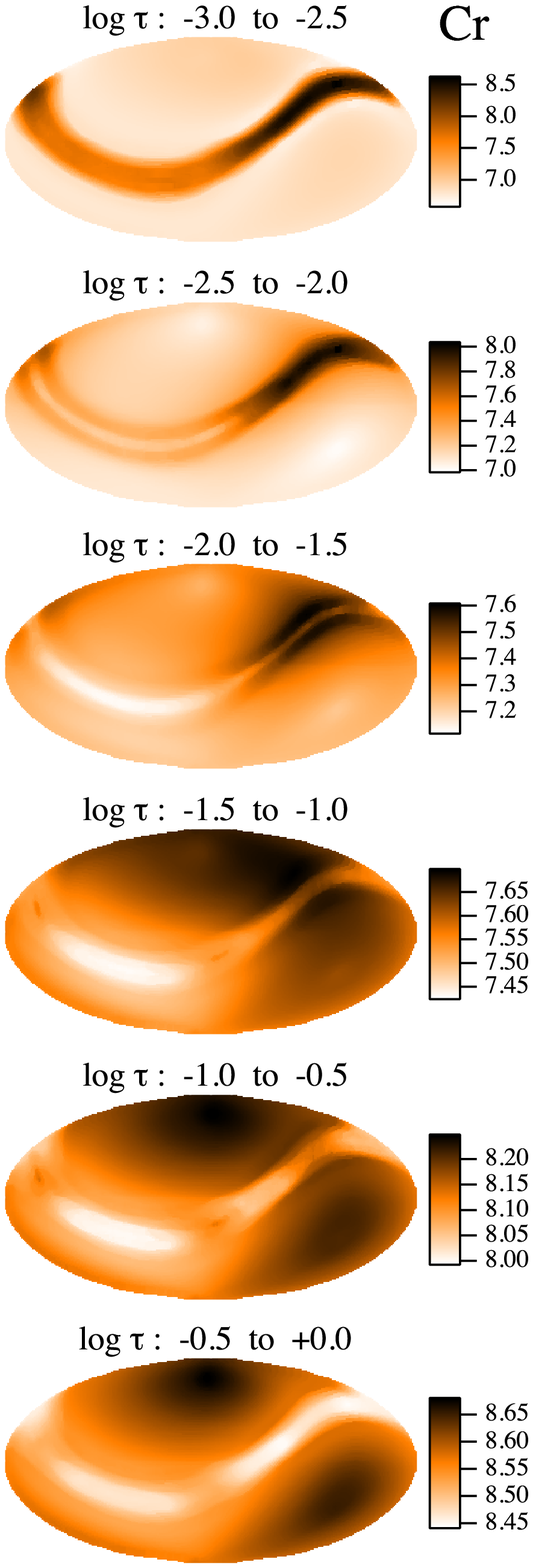}
\caption{
Tomographic view of the Cr abundance (Hammer equal-area projection). Six slabs corresponding to six contiguous optical depth ranges (indicated above each projection) are shown. Note that the relation between abundances and colour scale differs from one slab to the other. The solar abundance of Cr is 5.67~.}
\label{fig:ScansCol_Cr}
\end{figure}

\begin{figure}
\includegraphics[width=7.5cm]{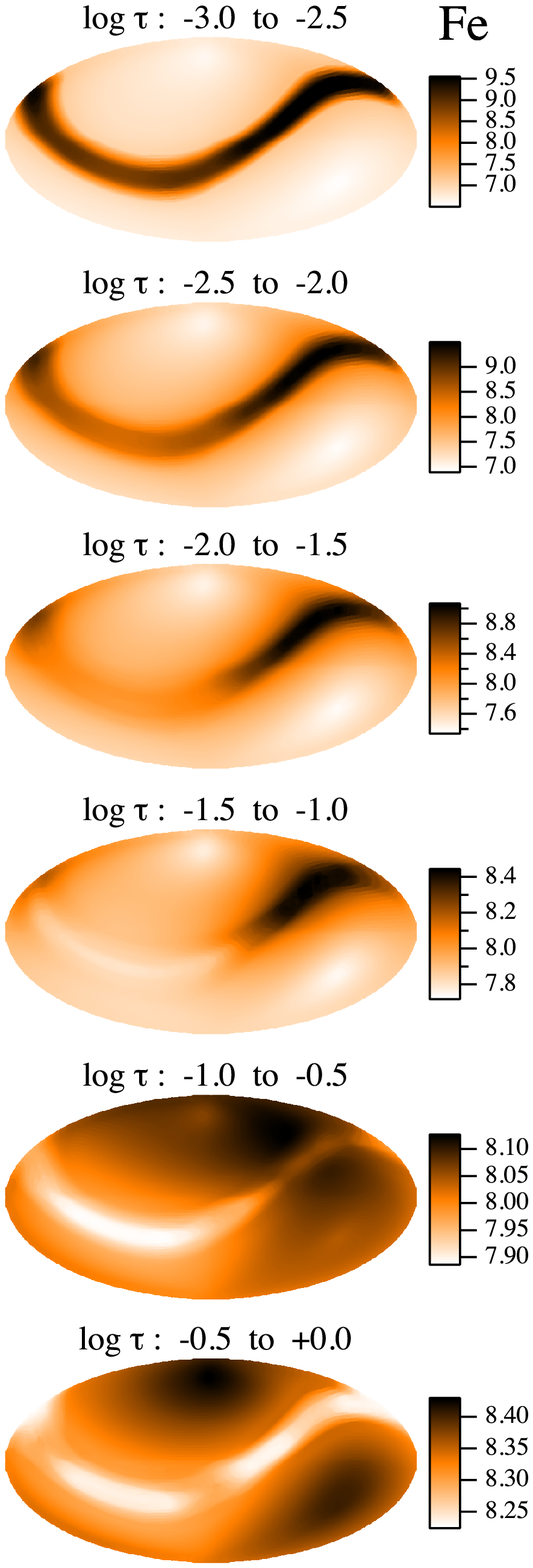}
\caption{
Same as Fig.~\ref{fig:ScansCol_Cr} for Fe. The solar abundance of Fe is 7.5~.}
\label{fig:ScansCol_Fe}
\end{figure}

In Fig.~\ref{fig:ScansCol_Cr} (chromium) and Fig.~\ref{fig:ScansCol_Fe} (iron) we show a tomographic view of the abundance distributions over the whole star as a function of depth. The 6 surface projections in these figures correspond to the abundances inside 6 slabs (i.e. 6 depths ranges) as defined in Sec.~\ref{sec:modeldistrib3D}. We have mentioned previously that in our grid, the models are composed of 72 layers ($-4.5\le\log{\tau}_{5000}\le 2.0$), but we have chosen to look only at 6 contiguous slabs in the usual line forming region ($-3.0\le\log{\tau}_{5000}\le 0.0$); each slab has an approximate optical thickness of 0.5\,dex. Indeed, as seen in Fig.~\ref{fig:ScansCol_Cr} and Fig.~\ref{fig:ScansCol_Fe}, 6 slabs provide sufficient resolution for visualisation of the vertical dimension. The optical depths (${\tau}_{5000}$) used to label the slabs are the ones taken from the model computed for solar homogeneous abundances (the first converged model in a run); this initial model is common to all the 81 models of the grid. We have verified that the ${\tau}_{5000}$ values of each layer of the final 81 equilibrium models differ by at most $\pm{0.2}$\,dex from those of the initial model.

As expected, the uppermost slab ($-3.0\le\log{\tau}_{5000}\le -2.5$) exhibits a finely shaped equatorial belt of overabundances for both elements. This is because the diffusion velocity is extremely sensitive to the magnetic field orientation at small optical depths. However, the abundances inside the rings are not uniform; for instance, even in the uppermost slab the overabundances in the left part of the plot are about 0.8\,dex lower than in the right part. For this reason, we will henceforth speak of a quasi-ring rather than of a ring. In addition, the abundance distributions change drastically as one goes deeper: the overabundant equatorial belt changes to large spots below $\log{\tau}_{5000}= -2.0$, parts of the belt become underabundant, and overabundances appear near the magnetic poles. We want to draw the attention of the reader to the fact that the relation between abundances and the colour scale differs from one slab to the other. The colour scale being the same for all the slabs, it may happen that from simple visual inspection the effects of abundance inhomogeneities on the emergent line profiles could be overestimated for the highest slabs. Indeed, the highest slabs contribute much less to the line profiles than the deepest slabs.

What makes the 3D maps shown in Figs.\,\ref{fig:ScansCol_Cr} and \ref{fig:ScansCol_Fe} so complex? At different depths, abundances are seen to vary over the star in different ways at different depths, giving rise to apparent abundance spots and pseudo-rings devoid of any symmetry. This is essentially due to the non-axisymmetric field structure assumed; the curve tracing the location of the horizontal field does not coincide with any curve of constant field strength, contrary to what happens in an axisymmetric field geometry. In the latter case, stratifications for a given magnetic latitude would be strictly the same for all magnetic longitudes, stratifications only change along the magnetic meridians.

\section{Discussion}
\label{sec:disc}

In order to achieve better predictions of 3D abundance stratifications in magnetic ApBp stars, we have computed a grid of 81 atmospheric models ($T_{\rm eff}=10\,000$\,K, $\log{g}=4.0$) with stratified abundances resulting from the simultaneous diffusion of 16 metals. These 81 models cover the range of magnetic strengths between 0\,G and 20\,kG, with a grid of magnetic inclination angles that reflects the sensitivity of the stratification profiles to the magnetic field. Based on this grid, we have modelled 3D distributions of chemical elements for a non-axisymmetric magnetic field geometry instead of the usual perfectly centred axisymmetric dipolar fields. In this paper, we have adopted a magnetic field geometry inspired by the published model of a real star (Fig.\,\ref{fig:Bcontour} and Fig.\,\ref{fig:BanglesContour}). Only two chemical elements (Cr and Fe, Fig.\,\ref{fig:ScansCol_Cr} and Fig.\,\ref{fig:ScansCol_Fe} respectively) are discussed. The results for 14 other chemical elements are available in our archives, but we do not deem it necessary to include them in this discussion. Our 3D computations could easily be extended to any other magnetic field geometry, provided that a map is available with modulus and orientation of the field vectors.

It is shown in Sec.\,\ref{sec:results} that above ${\log}{\tau}_{5000}=-2.5$ for Cr, and $-2.0$ for Fe, an overabundant quasi-ring exists in places where the magnetic lines are inclined by $90\pm{5}\degr$ with respect to the vertical. This is consistent with previous calculations for axisymmetric dipoles. In deeper layers however this simple structure clearly changes: the quasi-ring exhibits different vertical and horizontal extensions, depending on longitude and latitude. Parts of the quasi-ring may become underabundant, large structures or ``spots'' appear close to some portions of the quasi-ring and at the magnetic poles. It should be noted that rings do not necessarily exist deeper than ${\log}{\tau}_{5000}=-3$~ for all elements. We have found for instance that this is the case for Zn (not discussed in this paper). Most metals can develop such quasi-rings if one goes up high enough in the atmosphere. However, an overabundant ring, if formed above say ${\log}{\tau}_{5000}\approx{-5}$ would hardly affect line profiles (except for elements with very small solar abundance like rare earths, or Hg) and will remain undetectable with existing instruments.

For the moment, the existence of rings or quasi-rings as discussed above has only been established for equilibrium solutions to the diffusion problem -- but see the discussion in \citet{AlecianAl2015} for the limits of this hypothesis. We do not yet know if such structures also appear so clearly in time-dependent diffusion calculations \citep{AlecianAlStDo2011, StiftStAl2016}. It cannot be excluded that various parts of a ring or a quasi-ring appear on different timescales, making it potentially difficult to find a complete ring at a given age of the star.

Concerning our results for Cr and Fe, it is difficult to estimate the effect of the 3D abundance structure on the emergent line profile. The profiles will certainly be different from those expected for a globally constant stratification or for vertically constant but horizontally variable abundances; the effect will depend on atomic properties, wavelength, Zeeman pattern, depth of line formation ... Without extensive simulations it is impossible to predict how a technique like Zeeman Doppler mapping (ZDM) with its assumption of vertically constant abundances deals with this physical reality. In order to clarify the issue, we plan to extend the capabilities of {\sc{Cossam}} \citep{Stift2000, StiftStLeCo2012, Ramirez-VelezRaStNaetal2016y}, making it possible to approximate an ApBp atmosphere consisting of 7158 local atmospheres with individual elemental stratifications self-consistent with the local magnetic field, and to obtain full $IQUV$ Stokes spectra.

On the road towards improved modelling of element distributions in the atmospheres of main-sequence chemically peculiar stars, we plan to have a look again at time-dependent diffusion \citep{AlecianAlStDo2011, StiftStAl2016}, but now including the effect of mass-loss. Indeed, it is known that high mass-loss rates for $T_{\rm eff}\gta 16\,000$\,K do prevent atomic diffusion from stratifying abundances in atmospheres; they thus determine the upper limit in effective temperature of the ApBp star phenomenon (including non-magnetic atmospheres). On the other hand, diffusion models for cooler stars suggest that a mass loss of about $10^{-14}-10^{-13}$ solar mass per year acts in conjunction with atomic diffusion to yield the abundance anomalies observed in Am stars \citep{AlecianAl1996, VickViMiRietal2010}. Therefore mass-loss is certainly an important process to be included in numerical models. In magnetic atmospheres it is not unreasonable to assume anisotropic winds \citep{BabelBa1992r}, a scenario which can be expected to lead to some additional differences in the build-up of abundance stratifications in magnetic polar regions as compared to magnetic equatorial regions.

\section*{Acknowledgements}

All codes that have been used to compute the grid of models have been compiled
with the GNAT GPL Edition of the Ada compiler provided by AdaCore; this valuable
contribution to scientific computing is greatly appreciated. This work has been
supported by the Observatoire de Paris-Meudon in the framework of \emph{Actions
F\'ed\'eratrices Etoiles}. This work was partly performed using HPC resources
from GENCI-CINES (grants c2015045021, c2016045021). The authors want to thank
Dr. G{\"u}nther Wuchterl, head of the ``Verein Kuffner-Sternwarte'', for the
hospitality offered.




\bibliographystyle{mnras}
\bibliography{nondipole} 




%
%


\bsp	
\label{lastpage}
\end{document}